\documentclass[letter]{aa}  

\usepackage{lineno}
\usepackage{graphicx}
\usepackage{txfonts}
\AtBeginDocument{\mathcode`v=\varv}
\begin{document}

   \title{Backtracing the internal rotation history of the $\beta$~Cep star HD~129929}


   \author{S.J.A.J. Salmon
          \inst{1}
                    \and
          F.D. Moyano  \inst{1}          
          \and
           P. Eggenberger \inst{1}
                     \and
          L. Haemmerl\'e \inst{1}
          \and
          G. Buldgen \inst{1}
          }

   \institute{ Observatoire de Gen\`eve, Universit\'e de Gen\`eve, Ch. Pegasi 51, 1290 Sauverny, Switzerland \\
              \email{sebastien.salmon@unige.ch}
}
   \date{Received ;  }


 
  \abstract
   {HD~129929 is a slowly-rotating $\beta$ Cephei pulsator with a rich spectrum of detected oscillations, including two 
rotational 
multiplets. The asteroseismic interpretation  revealed the presence of radial differential rotation in this massive 
star of $\sim$9.35~M$_{\odot}$. The stellar core is indeed estimated to spin $\sim$3.6 times faster than the surface. 
The surface rotation was consequently derived as $v \sim 2$ km/s. This 
massive star represents an ideal counter-part to the wealth of space-based photometry results for main-sequence and 
evolved 
low-mass stars. Those latter have revealed a new, and often unexpected, picture of the angular momentum transport 
processes acting in stellar interiors. }
   {We investigate in a new way the constraints on the internal rotation of HD~129929, focusing on their interpretation 
for the evolution of the internal rotation during the main sequence of a massive star. We test separately hydrodynamic 
and magnetic instability transport processes of angular momentum.}
   {We used the best asteroseismic model obtained in an earlier work. We calibrated stellar 
models 
including rotation, with  different transport processes, to reproduce that reference model. We then looked whether one 
process is favoured to reproduce the rotation profile of HD~129929, based on the fit of the asteroseismic 
multiplets.}
   {The impact of the Tayler magnetic instability on the angular momentum transport predicts a ratio of the 
core-to-surface 
rotation rate of only 1.6, while the recently revised prescription of 
this mechanism predicts solid-body rotation. Both are too low in comparison with the asteroseismic inference. The models 
with only hydrodynamic processes are in good agreement with the asteroseismic measurements. Strikingly, we can also get a 
constraint on the profile of rotation on the zero age main sequence: likely, the ratio between the core and surface 
rotation was at 
least $\sim$1.7.}
   {Transport of angular momentum by the Tayler magnetic instability is discarded for this star. The models with 
pure hydrodynamical processes reproduce the asteroseismic constraints. This result is specific to a slow rotator and has 
to be verified more generally in other 
massive main-sequence stars. Constraints on the rotation in earlier stages of this star also offer new opportunity to 
test the impact of accretion during the pre-main sequence evolution.}
   \keywords{Stars: early-type --Asteroseismology -- Stars: rotation -- Stars: individual: HD 129929}

   \maketitle


\section{Introduction}

We have witnessed a leap in our knowledge of the internal rotation in intermediate- and low-mass stars 
thanks to space-based asteroseismology. At first, the determination of the core rotation rate in a large sample of red 
giants \citep{mosser12,gehan18} or precise radial profiles in subgiants and red giants 
\citep[e.g.][]{beck12,deheuvels12,deheuvels14,deheuvels15,triana17,dm18,fellay21}  has revealed values in strong 
disagreement with theoretical model predictions. It has pinpointed the need for additional mechanism(s) to extract 
efficiently angular momentum (AM) out of the core layers 
\citep[e.g.][]{eggenberger12,ceillier13,marques13,eggenberger17,eggenberger19a,deheuvels20}. We also got access to the 
rotation rates in the vicinity of the core layers of hundreds of $\gamma$ Dor pulsators 
\citep[e.g.][]{vanreeth16,ouazzani17,zwintz17,christophe18,li19,li20}, which are main-sequence progenitors of red giants. 
For a reduced sample of them, the rotation contrast between the core and envelope/surface was obtained, revealing almost 
rigid profiles of rotation \citep[e.g.][]{kurtz14,saio15,murphy16,li19,li20,saio21}. The preliminary work by 
\citet{ouazzani19} on these results also hints at the need for an additional transport process of AM during the main 
sequence. With results from other types of stars, we start having a view of the AM 
transport processes needed all across the Hertzsprung–Russell (HR) diagram \citep[see review by][]{aerts19}.


The $\beta$ Cephei pulsators are natural candidates for 
exploring transport mechanisms in more massive cases \citep[e.g.][]{bowman20,salmon22}. These 
main-sequence stars between $\sim$8 and 
25~M$_{\odot}$ present 
mixed-modes of oscillation probing the layers at the boundary of the stellar core, as well as 
pressure modes sounding the radiative envelope . They were 
among the first 
for which asteroseismology measured rotation rates in the stellar core and 
surface \citep[as of today, in five pulsators:][]{aerts03,briquet07,dziembowski08,desmet09,burssens21}. Too few or 
putative rotationally-associated non-axisymmetric modes were also reported in other $\beta$~Cep stars 
\citep[e.g.][]{mazumdar06,aerts06,briquet09,handler09,aerts11,briquet12}, which nevertheless makes them promising targets 
to increase the sample with known internal rotation. In the confirmed cases, except for $\theta$~Oph, clear 
indication of radial differential rotation was found for the other four pulsators, with 
core to surface 
ratios $\Omega_{\textrm{c}}/\Omega_{\textrm{S}}\sim$2 up to $\sim$8. 

In an attempt to contextualize these measurements, \citet{suarez09} showed in the case of $\nu$~Eri that between the 
two extreme hypotheses of rigid rotation and local conservation of AM, the latter was preferred to 
reproduce the asteroseismic data. 
We here focus on HD~129929, for which the implication of different AM transport processes was not 
investigated. Formally confirmed as a $\beta$~Cep star by \citet{waelkens83}, a dedicated campaign of 
observation \citep[hereafter A04]{aerts04} led to the detection of frequencies members of two multiplets, which offered 
to probe the internal rotation \citep{aerts03}. A detailed asteroseismic modelling of this star was given in 
\citet[][hereafter D04]{dupret04} who derived a core-to-surface rotation ratio of $\sim 3.6$, assuming a radial 
linear decrease of the rotation rate in the stellar interior. 

We have extended the analysis of D04, by focusing on the question of the internal rotation and its 
evolution during the main sequence in HD~129929. We used Geneva models including detailed treatment of rotation by 
hydrodynamic and magnetic instabilities, to determine which transport mechanisms lead to profiles of rotation 
in accordance with the one calibrated from asteroseismology.

We start by recalling in Sect.~\ref{section2} the properties of HD~129929 and presenting the physics of the 
different stellar models used in our analysis. In Sect.~\ref{section3} we show how the seismic constraints mark out the 
rotational profile of the star and its past evolution. We discuss in Sect.~\ref{section4} the implications 
of the determined rotation profile for different AM transport processes, before concluding.

\section{Dataset and methods}
\label{section2}

We report in Table~\ref{Table-Properties-Sect2} the observational frequency dataset determined by A04, and based on it, 
the properties of the best-fitting model found by D04. This asteroseismic 
modelling unambiguously identified the frequencies ($\nu$) as those of the radial 
fundamental 
pressure mode p$_1$\footnote{we use p$_i$ or g$_i$ to designate the $i$th radial order of a pressure or gravity mode, 
respectively, while $\ell_i$ stands for the angular degree $i$}; the complete 
p$_1$, $\ell_1$ 
mode triplet and two consecutive azimuthal-order components of the g$_1$, $\ell_2$ quintuplet. The 
frequency shifts between different azimuthal orders in multiplets depend on the rotation rate in 
specific regions of the star 
(see Sect.~\ref{section2-sub-rotsplit}). The triplet shows a small asymmetry, and we use its averaged splitting
value (as in D04), $\Delta_{\textrm{p}_1}$= 0.012130 c/d. The other splitting is 
$\Delta_{\textrm{g}_1}$=0.012109 c/d. The 
observational accuracy error on the frequencies was estimated to $\sim$ 10$^{-6}$ c/d.

\begin{table}[!]
\caption{Mode frequencies detected in HD 129929 (from A04) and parameters of the 
best stellar model reproducing this dataset (from D04).}             
\label{Table-Properties-Sect2}      
\begin{center}
                  \begin{tabular}{ll}        

\hline\hline                 
Seismic frequencies [c/d]  & Asteroseismic model \\
\hline
$\nu$ (p$_1$, $\ell_0$)= 6.590940 & M=9.35 M$_{\odot}$ ; $\log g$=3.905 \\ 
$\nu$ (p$_1$, $\ell_1$)= 6.966172 -- 6.978305 -- & T$_{\textrm{eff}}$=22 392 K ; X$_{\textrm{c}}$=0.353 \\
 6.990431 & X$_{\textrm{i}}$=0.7 ; Z$_{\textrm{i}}$=0.0188\\
$\nu$ (g$_1$, $\ell_2$)= 6.449590 -- 6.461699  &  $\alpha_{\textrm{ov}}$= 0.1 \\
 \hline     


\end{tabular}
\end{center}
{\flushleft T$_{\textrm{eff}}$ is the effective temperature ; $\log g$ the surface gravity ; 
X$_{\textrm{c}}$, the central H abundance ;  X$_{\textrm{i}}$ and Z$_{\textrm{i}}$ the initial H and metal mass 
fractions}; $\alpha_{\textrm{ov}}$ the overshooting parameter
\end{table}

\subsection{Stellar models}
\label{section2-sub-models}

We did not perform a new asteroseismic modelling of the star since the results by D04 were 
adequately reproducing the observed frequencies. We thus adopted the properties of their best stellar 
model. The best-fit asteroseismic model found by the authors revealed that HD~129929 is most likely 
a $\sim$ 9.35 M$_{\odot}$ 
star in the middle of its main-sequence hydrogen-burning phase. We adopted those stellar parameters (see 
Table~\ref{Table-Properties-Sect2}) to recompute a representative 
stellar model of HD~129929. This was done with 
the same stellar evolution code, CLES \citep{cles}, and following the same input physics than in D04 (see details in 
their paper). We assumed full redistribution or local 
conservation of AM, and verified whether they could predict an internal rotation profile reproducing the rotational 
splittings 
of the observed frequencies.

In a second step, we aimed at interpreting the asteroseismic constraints on the rotation profile in terms of AM transport 
mechanisms possibly at work in massive stars. We hence computed models including a coherent treatment of the 
rotation, with mean of the Geneva stellar evolution code \citep[GENEC,][]{genevacode}. The code relies on the assumption 
of shellular rotation as 
developed in \citet{zahn92}. It takes into account the meridional currents and shear instability for the estimation of 
AM transport along the evolution of the stellar models. It can also account for the impact of magnetic 
instabilities, following the Tayler-Spruit dynamo \citep{spruit02}, or its revision proposed by 
\citet{fuller19}. The advecto-diffusive equation for transport of AM in radiative zone(s) while accounting for the 
aforementioned processes reads as:

\begin{equation}
 \rho \frac{d}{dt} (r^2\Omega) = \frac{1}{5r^2}\frac{\partial}{\partial r}(\rho r^4 \Omega U)+ 
\frac{1}{r^2}\frac{\partial}{\partial r}\left[ (D_{\textrm{shear}}+\nu_{\textrm{M}}) \rho r^4 \frac{\partial 
\Omega}{\partial r} \right],
\end{equation}
where $r$ is the radius, and $\rho$, $\Omega$ the mean density and mean angular velocity on an isobar, 
respectively. The radial component of the meridional circulation is given by $U$. The diffusion coefficient 
$D_{\textrm{shear}}$ is the one for the transport of AM by the shear instability \citep[following][]{maeder97}, 
while $\nu_{\textrm{M}}$ represents the diffusion coefficient from magnetic instabilities. When no magnetic instability 
is considered, $\nu_{\textrm{M}}=0$ and the models then computed are referred to as pure hydrodynamical. The magnetic 
instabilities accounting for the transport of AM were computed following first the original prescription of the
Tayler-Spruit dynamo. In this case, $\nu_{\textrm{M}}$ takes the form of the $\nu_{\textrm{TS}}$ coefficient as given 
in \citet[][see their Eq. 1-2]{eggenbergermag}. As recalled by these authors, to whom we refer for details, this 
mechanism 
works in regions presenting shear and is activated above a triggering threshold, but it is inhibited by chemical 
gradients. We also 
considered the Fuller revision of this mechanism, in which case 
$\nu_{\textrm{M}}$ takes the form of $\nu_{\textrm{T}}$ as given in Eq.~3 of \citet{eggenbergermag}. These Geneva 
models were constructed to reproduce the stellar properties of the asteroseismic model as recalled in 
Table~\ref{Table-Properties-Sect2}. 

\subsection{Rotational Splittings}
\label{section2-sub-rotsplit}

The frequency 
separation of different azimuthal-order modes in a multiplet is directly related to the rotation rate in the 
layers where these 
modes propagate. For stars with rotation frequencies much smaller than the oscillation frequencies, a simple perturbation 
approach 
can be used. We only considered a first-order expression , since 
HD~129929 is a clear slow rotator, $v \sin i \lesssim 13$ km/s: in comparison, stars with similar 
spectral types present projected velocities on average between 160 and 200 km/s \citep[e.g.][]{royer09}. The first order 
accounts only for Coriolis effects, while higher-order terms also 
accounts for 
centrifugal effects \citep[see e.g.][]{goupil11}. As mentioned earlier, here the $\Delta_{\textrm{p}_1}$  
splitting presents a small asymmetry ($\sim 7 \times 10^{-6}$ c/d). Although such a feature is expected from higher-order 
effects of rotation, the rotation velocity of the star is so low that these effects are likely to be negligible. This 
is in line with computations up to the 3rd order for a typical $\beta$ Cep model at various rotation rates 
presented in \citet{ouazzani12}. Limiting to the 1st order, as demonstrated by \citet{ledoux51}, the 
shift in the frequency of a non-axisymmetric mode then reads as:
\begin{equation}
\label{eq-1}
 \nu_{n,\ell,m}=\nu_{n,\ell,0} + m \int_0^R K_{n,\ell} \Omega \ dr,
\end{equation}
where $\nu$ is the frequency of a ($n,\ell,m$) mode, $n$ the radial order, $m$ the azimuthal order, $R$ the stellar 
radius, and $K_{n,\ell}$, the 
rotational kernel of the mode. The latter is built from the mode eigenfunctions, and has to be computed from a stellar 
structure model. Hence, it must be representative of the star under investigation. This is why we recomputed the 
asteroseismic solution of D04; from this stellar model, we derived the rotational kernels of the p$_1$,$\ell_1$ and 
g$_1$,$\ell_2$ modes with the LOSC adiabatic oscillation code, in its standard setting \citep[see details 
in][]{losc}.

We finally tested the ability of various rotation profiles $\Omega (r)$ to reproduce the observed values of  
$\Delta_{\textrm{p}_1}$ and $\Delta_{\textrm{g}_1}$ by comparing the theoretical counterparts of these splittings, 
evaluated with Eq.~\ref{eq-1}, and using the simple following merit function:
\begin{equation}
 \chi^2=\frac{1}{2}\sum_{i=1}^{2} 
\frac{(\Delta_{\mathrm{obs,i}}-\Delta_{\mathrm{th,i}})^2}{\sigma_{i}^2},
\label{eq1-section2}
\end{equation}
with $\sigma_{i}^2$ the observational error ($\sigma_{i}$ is set to 10$^{-6}$ c/d).

\section{Analysis of the internal rotation constraints}
\label{section3}
\begin{figure}
\centering
\includegraphics[width=0.5\textwidth]{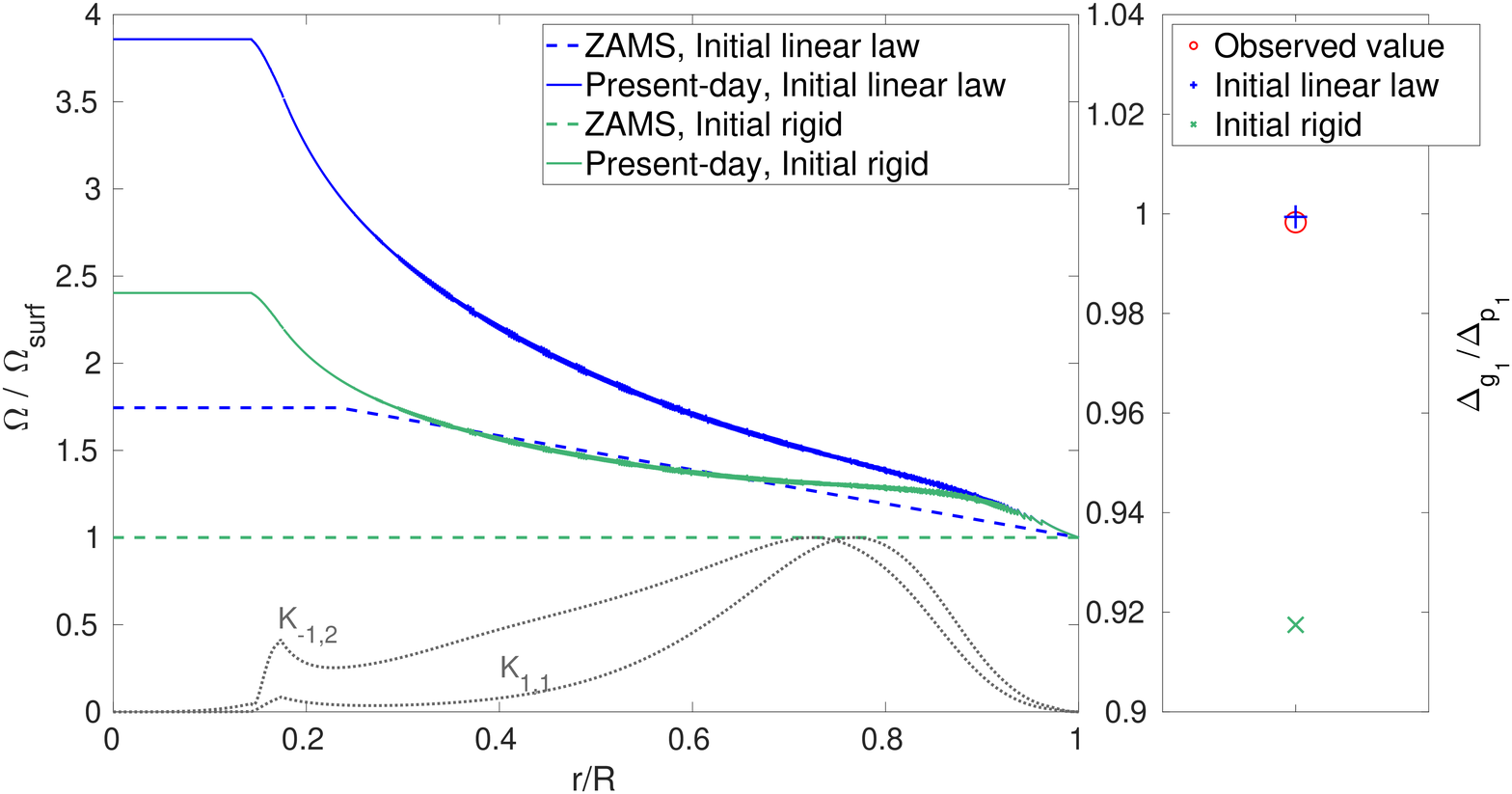}
      \caption{Left panel: normalised profiles of rotation in the CLES model of HD~129929 (solid lines) as a 
function of the 
normalised radius. The 
profiles are those that best fit the observed rotational splittings assuming a rigid (green) or linear (blue) law 
on the ZAMS and local AM conservation during evolution. The related ZAMS profiles (corresponding to the minimums in 
$\chi^2$ shown in Fig.~\ref{Fig2-ChiS}) are shown in dashed lines. The normalised rotational 
kernels of the p$_1$ and g$_1$ modes are represented by the grey dotted lines.
Right panel: comparison of the splitting ratio between the observed value (red) and, correspondingly to the left panel, 
the best-fit solutions assuming a rigid (blue) or linear (green) profile of rotation on the ZAMS.}
         \label{Fig1_Profiles_Seismo}
\end{figure}

We show in Fig.~\ref{Fig1_Profiles_Seismo} as dotted lines the kernels $K_{1,1}$ and $K_{-1,2}$, respectively associated 
with the observed splittings $\Delta_{\textrm{p}_1}$ and $\Delta_{\textrm{g}_1}$. Once computed the 
integrals of the former, it is straightforward that the splittings cannot be reproduced with a rigid rotation profile, as 
stated previously in \citet{aerts03}. This discards the assumption of full redistribution of AM in the star 
during its evolution.

\begin{figure*}[!]
\centering
\includegraphics[width=0.48\textwidth]{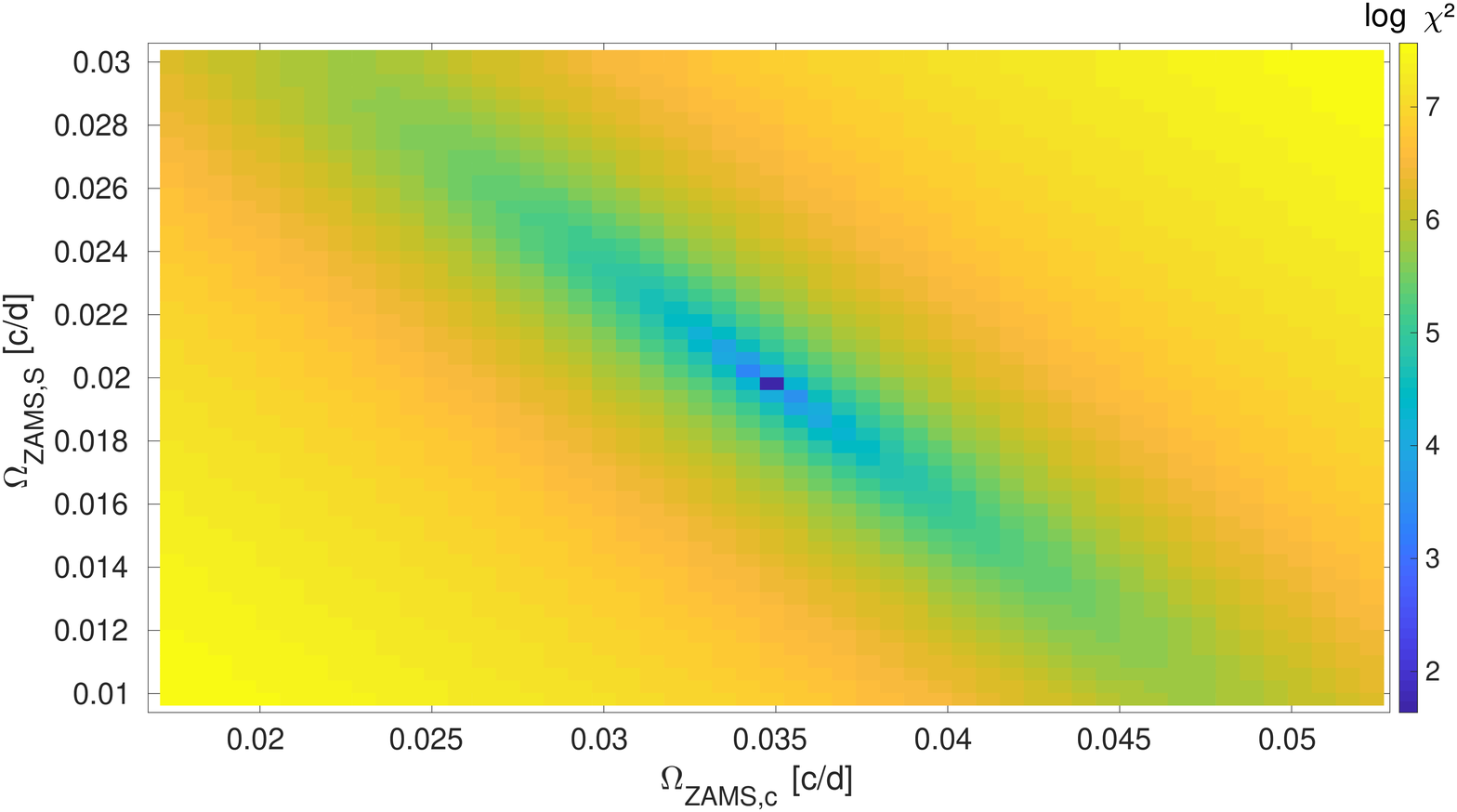}\includegraphics[width=0.48\textwidth]{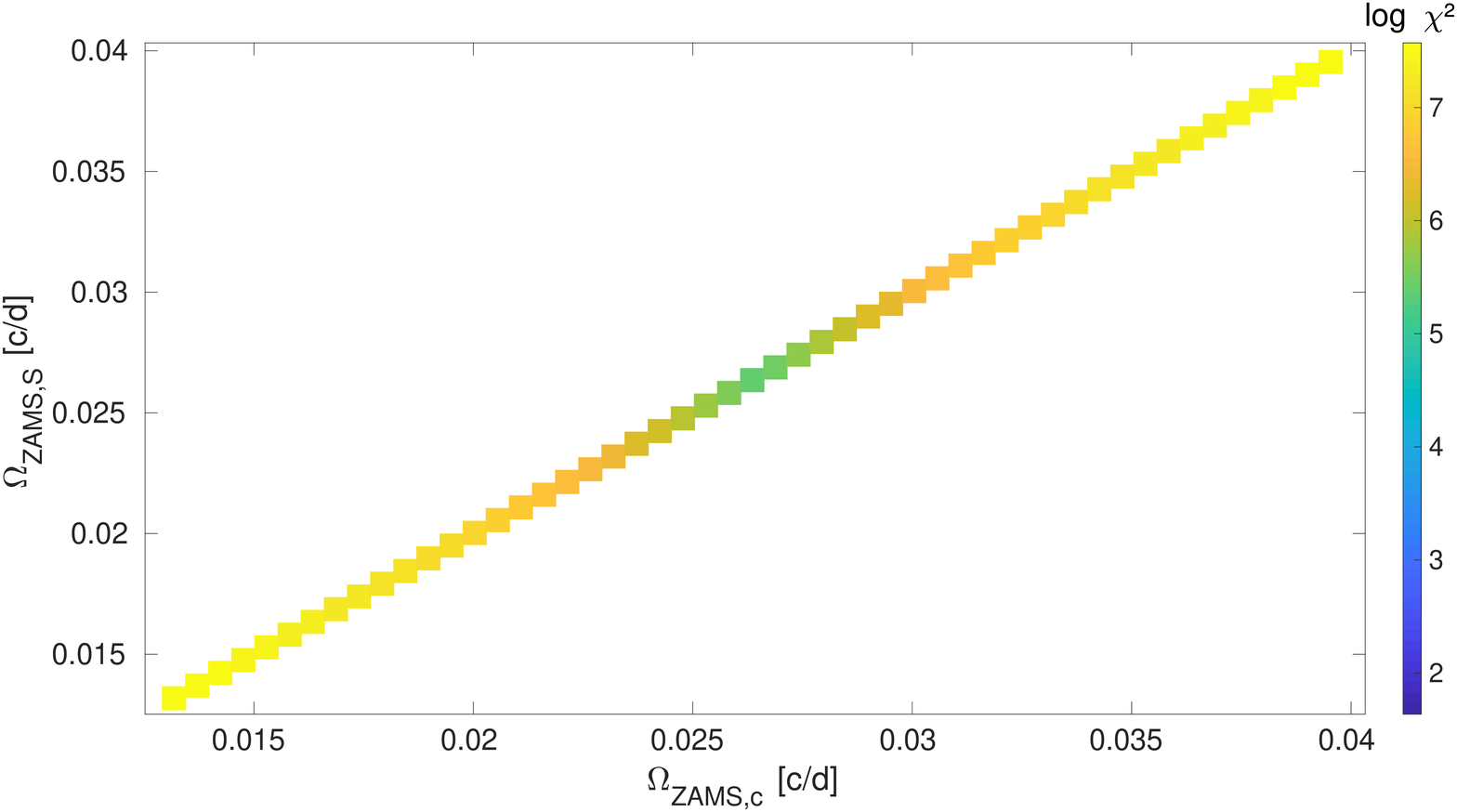}
\caption{Map of the $\chi^2$ values based on the comparison of HD~129929 observed and theoretical rotational 
splittings, as in Eq.~\ref{eq1-section2}, assuming a linear (left panel) or rigid (right panel) profile of 
rotation on 
the ZAMS, and local AM conservation along the evolution to the current stage of the star.}
\label{Fig2-ChiS}
\end{figure*}

We then assumed the opposite with no redistribution of AM by including local conservation alongside 
the stellar evolution.  We first considered a case where a rigid profile is assumed on the zero age main sequence (ZAMS). 
We found that once evolved to the current stage of HD 129299, the minimum value in 
$\chi^2$ is very large, 2.64 $\times 10^5$, as depicted in the right panel of Fig.~\ref{Fig2-ChiS}. The rotation 
profiles corresponding to this solution are shown in green in the left panel of Fig.~\ref{Fig1_Profiles_Seismo}, both at 
the ZAMS and once at the current age of HD~129929. We notice very small wiggles (same for the blue curve) 
of the profile in the envelope, which are here resulting from the number of points describing the model 
and the numerical integration scheme used to follow the AM conservation along the stellar track, without impact 
on the 
computed splittings. As indicated by the large $\chi^2$ value, the ratio between the core 
and surface rotation rates of this green profile is too small and cannot reproduce the splittings. We 
represented in the 
right panel of this figure the ratio of the splittings ($\Delta_{g_1}/\Delta_{p_1}$), which is here an estimate of the 
rotation contrast between the core and envelope. The splitting ratio of this model with a rigid 
rotation on the ZAMS is 
clearly lower than the observed value, confirming it has a too low rotation contrast between its core and surface.

However, during their early stages of evolution, stars of $\sim$9~M$_{\odot}$ are fully radiative 
during a large fraction of the accretion 
phase \citep{haemmerle19}. Once the star reaches the ZAMS, most of the radiative layers are as such since they 
were accreted, avoiding convective AM transport. In these short timescales of the pre-main sequence phase (a few 
10$^5$ years for a 9~M$_{\odot}$), meridional 
circulation and shear diffusion are expected to remain negligible, in particular for slow rotators, so that each radiative 
layer keeps the AM it advected at accretion as the star contracts, and the rotation 
profile reflects essentially the AM accretion history \citep{haemmerle17}, given by the rotational properties of the 
pre-stellar cloud. We thus assumed a non-rigid profile on the ZAMS, with rigid rotation in 
the convective core, and a simple linear relation in the radiative envelope. Letting evolve the model at the age of 
HD~129929, we succeed to 
find a rotation profile that reproduces the splittings. The solution is well defined and non-degenerate, 
as shown in the left panel of Fig.~\ref{Fig2-ChiS}. Starting with an initial 
$\Omega_{\textrm{ZAMS,c}}/\Omega_{\textrm{ZAMS,S}}$\footnote{$\Omega_{\textrm{ZAMS,C}}$ and $\Omega_{\textrm{ZAMS,S}}$  
are respectively
the core and surface rotation rates at the ZAMS} of 1.74, the profile 
evolves to a present-day value of 3.86. In the right panel of Fig.~\ref{Fig1_Profiles_Seismo}, this 
model reproduces the observed splitting ratio, confirming that its profile of rotation (blue line 
in the left panel) presents an adequate contrast between the core and surface. This is in good agreement with the 
simpler linear model of D04, which revealed a similar contrast between the core and surface. The 
freedom of varying the rotation profile to reproduce the splittings is actually limited in this case, as a result of the 
two kernels we have at our 
disposal. More specifically, 
$K_{-1,2}$ (see Fig.~\ref{Fig1_Profiles_Seismo}) is clearly associated with a mixed mode. While it bears all the 
information we have on the central layers, it nevertheless is sensitive to a large extent on the envelope and surface 
layers, due to its larger amplitude in these regions.

\section{Angular momentum transport processes}
\label{section4}

Assuming local AM conservation throughout the MS evolution yields a realistic 
profile of rotation able to reproduce the observational splittings, but also unambiguously gives a picture of it at the 
ZAMS. We 
thus compared this solution to more sophisticated stellar models that 
treat AM transport with physically-motivated mechanisms. This was done with 9.35~M$_{\odot}$ Geneva models as 
explained in Sect.~\ref{section2-sub-models}. The linear rotation profile on the ZAMS we determined in 
Sect.\ref{section3} was adopted as an initial condition for these refined models. The models were calibrated to 
reproduce the same location in the HR diagram as the CLES model of HD~129929. We compared their structure and 
found that the quantities relevant for asteroseismic properties were very similar between CLES and GENEC 
models.

\begin{figure}[!]
\centering
\includegraphics[width=0.5\textwidth]{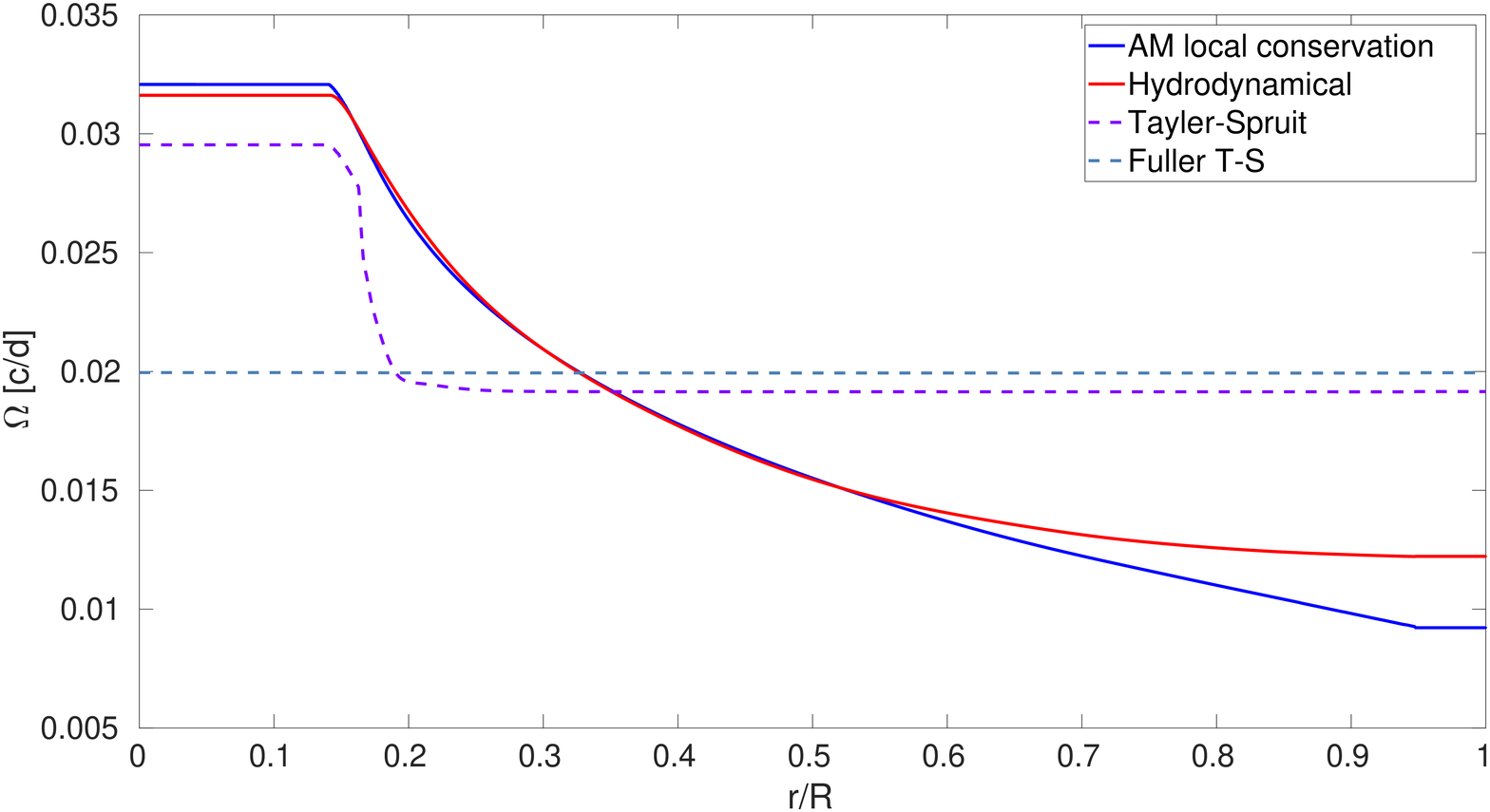}
\caption{Profiles of rotation as a function of the normalised radius for the different Geneva models of 
HD~129929. The different mechanisms of AM transport included in each model are indicated in the legend. Dotted lines 
distinguish those with magnetic instabilities.}
\label{Fig3}
\end{figure}

The profiles of rotation resulting from the different transport mechanisms used through the evolution of the models are 
shown in Fig.\ref{Fig3}. The hydrodynamical model predicts a 
very similar profile as in the local conservation assumption. The meridional circulation is the main driver 
of AM transport through advection at large scales. However, we deal in the present case with a slow rotator, and the 
circulation is consequently very weak, resulting in a limited transport of AM. Since the hydrodynamical model is very 
close to the one with AM local conservation, the profile of rotation it predicts is in accordance with the asteroseismic 
constraints for this star.

We also represent the two profiles of rotation obtained with the magnetic instabilities, either following the 
original prescription of the Tayler-Spruit dynamo or the Fuller revision. In the former case, differential rotation can 
however develops close to the convective core where the chemical gradient weakens the instability, later on allowing 
for an efficient transport 
of AM. Yet, the contrast between centre and surface cannot account for the asteroseismic constraints. Furthermore, the 
contrast developed by this  mechanism is relatively insensitive to the initial conditions. In the latter case,
the AM transport is much more efficient and leads to rigid rotation, which is clearly discarded 
in this star. The coupling between central and superficial layers in these two models with magnetic instabilities are
both too strong: thus, the asteroseismic observations in the case of the HD~129929 star do not favour such instabilities.

\section{Conclusion}
\label{conclusion}

Starting from the well-established asteroseismic solution for the $\beta$~Cep star HD~129929, we have studied in 
detail the consequences of its asteroseismic properties on the internal rotation of the star, including its past 
evolution on the main sequence. Assuming local conservation of angular momentum during the main sequence, we 
have determined a non-degenerate solution showing that the present-day rotation at the centre of the star is probably 
$\sim$ 3.86 times greater than at the surface. This result is in good agreement with the solution found by D04, which was 
a factor 3.6. This latter relies on the approximation of a linear decreasing law of the rotation rate from the convective 
core to the surface, without consideration for the past evolution of the star.

Given the non-degeneracy of the solution obtained under local conservation of angular momentum, we could backtrace 
the 
evolution of the internal rotation of this star. Assuming a linear law for the profile of rotation on the zero age main 
sequence, we have determined that the core of the star was then rotating at least $\sim$1.74 times faster than the 
surface. 
This results highlights the potential for asteroseismology not only for constraining the current structure of 
a massive star but also conditions prevailing in its early stages of evolution. The limits given on the internal rotation 
at the ZAMS are offering a new testbed for conditions of formation of these stars and their 
pre main-sequence evolution, and for example, how angular momentum redistributes during episodes of accretion 
\citep{haemmerle17}.

In a next step, we have compared our asteroseismic solution to the rotational profiles predicted by models representative 
of HD~129929, including different transport processes. First 
considering the sole 
combination of meridional circulation and shear instability, we found it leads to a very low degree of redistribution 
of the angular momentum: the meridional circulation is hindered to act efficiently because the star is a slow rotator 
(our asteroseismic solution predicts a surface velocity of 2.29~km/s). 
It leads to a core-to-surface ratio fully compatible with the asteroseismic profile deduced in the star. Models which 
included magnetic instabilities, either the Tayler-Spruit dynamo or its Fuller revision, show a strong coupling between 
the core and envelope. As a result there is an efficient redistribution of angular momentum that flattens the profiles of 
rotation, which is clearly rejected by the asteroseismic constraints in this case.

While magnetic instabilities are good candidates for explaining the profile of rotation in the Sun 
\citep{eggenbergermag} or partially low-mass stars in evolved stages \citep[e.g.][]{fuller19}, HD~129929 is a 
first proven example of their inadequacy in a main-sequence massive star. To the contrary, the pure hydrodynamical case 
appears a good 
candidate to explain the rotational properties of this star. This is 
in contrast to the conclusions 
for low- and intermediate-mass stars, in which additional transport processes are 
expected \citep[e.g.][]{ouazzani19,eggenberger19a}. The present case remains specific as HD~129929 is a slow rotator. In 
the wake of the present effort, the 
other $\beta$~Cep stars for which we have rotational 
splittings will allow us to further constrain the transport processes at work in massive stars.

%

\begin{acknowledgements}
We are grateful to G. Meynet and M.-A. Dupret for helpful discussions and remarks on this work.
S.J.A.J.S., F.D.M., P.E. and L.H. have received funding from the European Research
Council (ERC) under the European Union’s Horizon 2020 research and innovation program (grant agreement No 833925, 
project STAREX). G.B. acknowledges funding from the SNF AMBIZIONE grant No. 185805 (Seismic inversions
and modelling of transport processes in stars).
\end{acknowledgements}

\bibliographystyle{aa}
\bibliography{biblio}

\end{document}